\documentclass[aps,prl,reprint,floatfix,longbibliography,twocolumn]{revtex4-2}
\usepackage{amsmath,amssymb,amsthm,hyperref,graphicx}
\hypersetup{hidelinks}
\usepackage[capitalise]{cleveref}
\hypersetup{colorlinks=true, linkcolor=blue, citecolor=red, urlcolor=blue}
\begin{document}
\title{Harnessing Plasmonic Heating For Switching In Antiferromagnets}
\author{H. Y. Yuan$^{1}$}
\email{Contact author: hyyuan@zju.edu.cn}
\author{Yizheng Wu$^{2}$}
%\affiliation{$^{1}$Institute for Advanced Study in Physics, Zhejiang University, 310027 Hangzhou, China}
\author{Olena Gomonay$^{3}$}
%\author{Bert Koopmans$^{2}$(?)}
\affiliation{$^{1}$Institute for Advanced Study in Physics, Zhejiang University, 310027 Hangzhou, China}
\affiliation{$^{2}$Department of Physics and State Key Laboratory of Surface Physics, Fudan University, Shanghai 200433, China}
\affiliation{$^{3}$Institute of Physics, Johannes Gotenberg-University Mainz, 55099 Mainz, Germany}

\date{\today}

\begin{abstract}
Heat waste is a bottleneck in the development of green information technologies and much effort has been devoted to suppress the heating effect in both electronic and spintronic devices. Here we take an alternative approach and show that controllable heating at the nanoscale can actually benefit information processing. In particular, we study a hybrid nanostructure consisting of a metallic square frame and an antiferromagnetic (AFM) thin film and show that the plasmonic heating can reversibly switch two perpendicularly-oriented AFM domains without the assistance of magnetic fields and electric currents. The required switching energy is at the order 1 nJ, three to six orders of magnitude lower than the current-driven AFM switching. The physical mechanism arises from the thermal-induced strain fields inside the frame, which couple to and manipulate the magnetic orientation via magnetoelastic effect. The strain field direction can be well controlled by selectively exciting the longitudinal and transverse plasmon modes by varying the polarization of the waves, which readily allows for a reversible switching of the AFM vector. Our findings provide tremendous opportunities for optically manipulating the magnetism with ultralow energy consumption and may further promote the interdisciplinary study of photonics, acoustics and spintronics.
\end{abstract}

% subterahertz 0.1-1THz
\maketitle
%\section{Introduction}
\textit{Introduction.--}
With the rapid accumulation of data in modern society, it is imperative to find efficient and sustainable ways for information processing. Traditional semiconductor devices face fundamental limitations due to the unavoidable heating effects generated by electron transport. Spintronics that manipulates the spin degree of freedom of electrons in condensed matter structures can, in principle, suppress the charge flow; however, the heating effect is usually not negligible in practical implementations \cite{Ralph2008,Manchon2019,Gobel2021}. 
Among various spintronic platforms, antiferromagnets (AFM) stand out as promising candidates for next-generation spintronic devices owing to their ultrafast dynamics and robustness against stray fields. However, electric manipulation of AFM domains generally relies on larger current densities, where Joule heating becomes not only a by-product but in many cases an indispensable contributor to the switching mechanisms \cite{Chiangprl2019,Zhangprl2019,Baldratiprl2020,Meernl2021,Wuprb2024}. Thus, efficient control of AFM domains remains an outstanding challenge. 

To mitigate excessive heating, conventional approaches typically explore alternative physical mechanisms for magnetic control, including the voltage-based method \cite{FertRMP2024}, optical and acoustic techniques \cite{YangAPR2021} and purely magnonic means \cite{ChumakNP2015,YuanQM}.  Alternatively, rather than suppressing heating, a different and largely unexplored paradigm is to directly exploit nanoscale heating as a resource for useful applications. Thermoplasmonics, which exploits the collective electron oscillations in metallic structures so-called plasmons, offers precisely such an opportunity. Plasmons can confine the electric fields in sub-wavelength scale and generate highly structured temperature profiles at the nanoscale, which has found applications in bio-imaging, medical therapy, and nanochemistry \cite{Baffou2013,Jauffred2019,Cui2023,Maier2015}. Whether and how we can harness the plasmon-induced thermal landscape to control magnetism remains an open and compelling question.

%One prominent example is the spin Seebeck effect \cite{Uchida2010,Bauer2012}, where a temperature gradient can generate a spin current that can be converted into the electric voltage, which is interesting for both fundamental physics study and potential applications as thermoelectric devices. Building on this concept, we pose a compelling question: Can the heating effect be exploited to directly manipulate the magnetization direction instead of the magnetic excitations? This outlines the motivation for our investigation. 
%

%Recently, there have been some hints that the Joule heating effect may be important during the spin-torque driven magnetic switching \cite{Chiangprl2019,Zhangprl2019,Baldratiprl2020,Meernl2021,Wuprb2024}. Yet to be known is whether we can deliberately control and manipulate the heat current at the nanoscale. While a homogeneous heating across a magnet only reduces the crystalline anisotropy and saturation magnetization without exerting a net spin-torque to switch the magnetization, an inhomogeneous heating can directly couple to the magnetic vector through magnetoelastic interaction and act as an effective field to control the magnetization direction. Notably, it is known to the community of nanophotonics that surface plasmons in metallic nanoparticles provide a natural way to control the heating at the nanoscale, which has found interesting applications in . 

\begin{figure}
	\centering
	\includegraphics[width=0.45\textwidth]{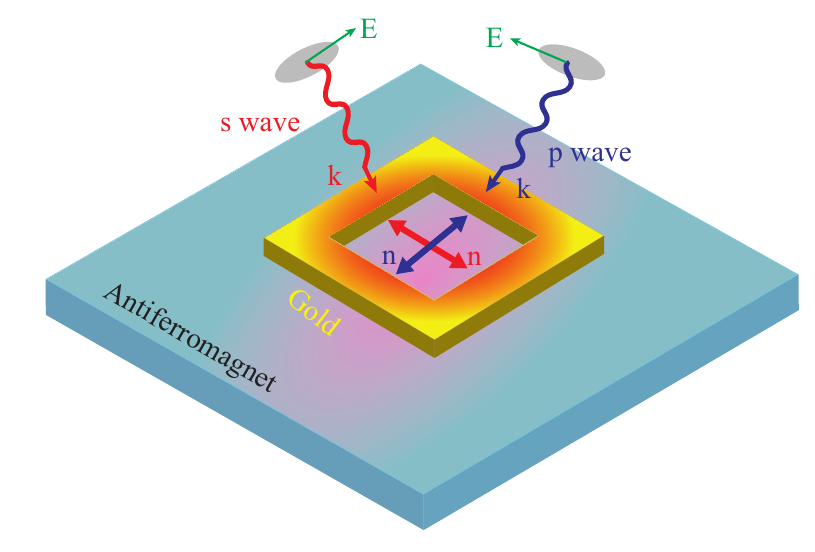}\\
	\caption{Schematic illustration of a gold nanoframe and magnetic thin film hybrid structure. The antiferromagnetic vector within the frame can be reversibly switched by alternating the polarizations of the incident light, enabled by plasmon-induced selective heating and strain effects.}\label{fig1}
\end{figure}

In this Letter, we address this question by examining a simple hybrid nanostructure consisting of a gold nanoframe and an AFM insulator, as shown in Fig. \ref{fig1}. We demonstrate that the perpendicularly-oriented AFM domains inside the frame region can be switched reversibly by alternating the polarization of the incident optical waves. Remarkably, the energy consumption of a single switching is three to six orders of magnitude lower than that needed for the electric-current-driven switching. The underlying mechanism can be understood from the interplay of plasmons, inhomogeneous heating and the resulting strain fields. Optical excitation of the nanoframe produces strongly anisotropic temperature gradients that drive the directional thermal strains, which couple to the AFM vector through magnetoelastic interactions. By choosing the polarization of the incident light, we selectively excite longitudinal or transverse plasmon modes in different arms of the frame, thereby controlling the sign of the induced strain and enabling reversible switching between two orthogonal AFM domains. Our results reveal a conceptually distinct route to magnetic control in which nanoscale heating--rather than being mitigated--is intentionally engineered to manipulate the AFM vector.
The thermoplasmonic mechanism provides an all-optical and low-energy-consuming route to manipulate antiferromagnetism and opens opportunities at the interface of spintronics, photonics and acoustics.

%When an optical wave shines on the nanoframe, only the two arms that align with the direction of the electric field component of the wave are strongly excited, generating plasmons inside the arms. These plasmonic excitations serve as an energy source to heat the nanoframe and the magnetic substrate, generating an inhomogeneous temperature distribution inside the frame region. Then the subsequent nonuniform deformation of the magnetic lattice induces a strain field, acting as an effective magnetic field to switch the magnetic orientation. When the electric polarization of the optical waves changes, the other two frame arms are excited, generating an opposite strain field to switch the magnetic orientation back. 

\begin{figure}
	\centering
	\includegraphics[width=0.48\textwidth]{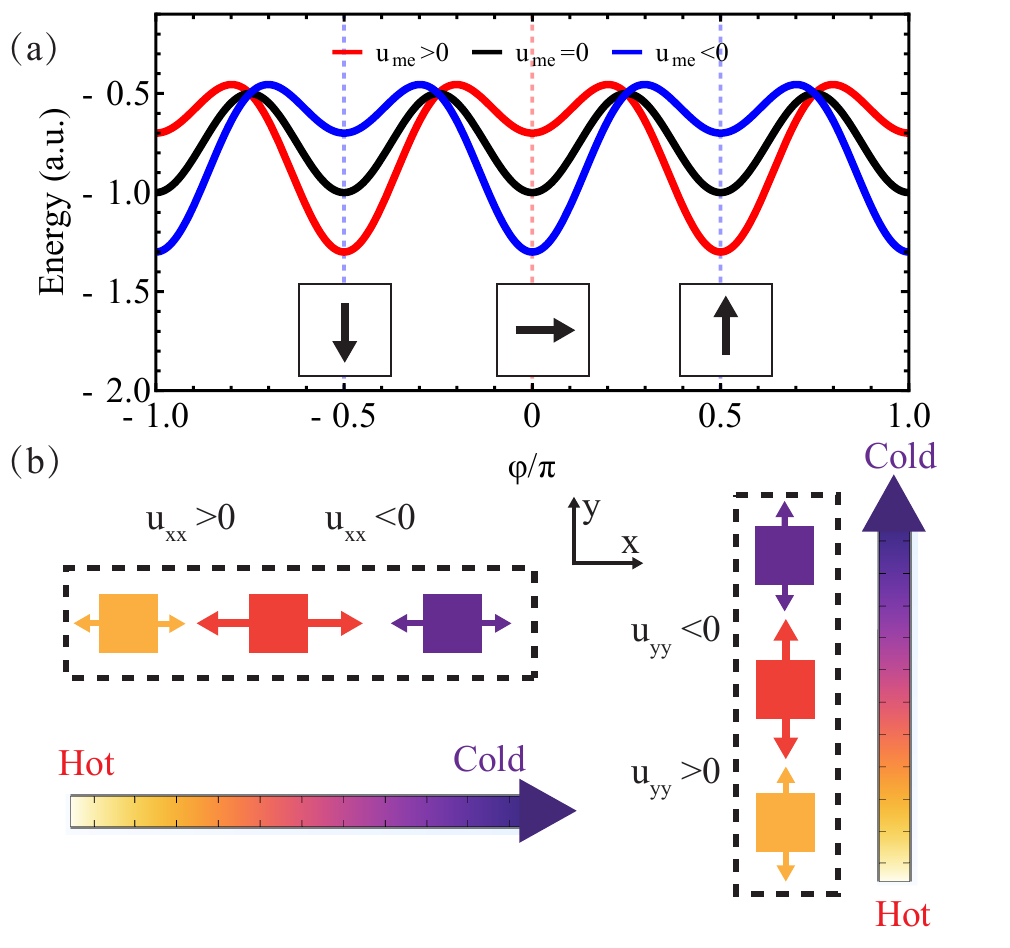}\\
	\caption{(a) Energy landscape of the antiferromagnet with and without the strain field $u_{me}$. The insets with arrows sketch the direction of the Néel vector. (b) Illustration of the lattice deformation and the corresponding strain field in an elastic medium under a horizontal and vertical temperature gradient, respectively.}\label{fig2}
\end{figure}

\textit{Strain-field-induced magnetic switching.--}
To illustrate the essential physics, we consider an AFM thin film NiO described by the Hamiltonian \cite{Meernl2021}
\begin{equation}\label{eqn1}
\begin{aligned}
\mathcal{H} &= 2\mu_0M_sH_{me}(n_x^2-n_y^2)(u_{xx}-u_{yy}) \\
&+ 2\mu_0M_s[H_{\parallel}n_z^2-H_{\perp}(n_x^4+n_y^4)],
\end{aligned}
\end{equation}
where $\mathbf{n}=(n_x, n_y,n_z)$ is the Néel vector or AFM vector that characterizes the magnetization difference of the two sublattices. The first and second terms represent the magnetoelastic energy and intrinsic magnetic anisotropy, respectively, with $\mu_0$, $M_s$, $H_{me}$, $H_{\parallel}$, $H_{\perp}$ being the vacuum permeability, saturation magnetization of each sublattice,
magnetoelastic field strength, out-of-plane and in-plane anisotropy fields, respectively. $u_{ij}=(\partial_j u_i + \partial_i u_j)/2$ is the strain tensor component with $u_i$ being the displacement vector along $i$-th direction ($i,j=x,y,z$) \cite{Landaubook,Hetnarski}. Since $H_\parallel$ imposes a strong easy-plane anisotropy in NiO, the Néel vector remains confined to the film plane, i.e. $n_x = \cos \varphi, n_y=\sin \varphi$, whereas $H_\perp$ determines the fourfold anisotropy within the plane. By varying the total energy $\mathcal{H}$ with respect to $\theta$ and performing a stability analysis, one can find that the Hamiltonian hosts four local minima: $n_x=\pm 1$ and $n_y=\pm 1$. 
In the absence of magnetoelastic field, i.e. $u_{me}\equiv u_{xx}-u_{yy}=0$, these four states are degenerate in energy. Once a nonzero $u_{me}$ is introduced, one pair of states remains stable while the other pair of states becomes metastable (red and blue lines in Fig. \ref{fig2}(a)). For $u_{me}>H_{\perp}/H_{me}$ ($u_{me}<-H_{\perp}/H_{me}$), only the states $n_y =\pm 1$ ($n_x=\pm1$) are stable. The critical value of $u_{me}$ for NiO is about $10^{-5}$ \cite{Meernl2021}. In principle, it is enough to remove the degeneracy between the two types of domains to induce switching, where the strain field can help to overcome the pinning barrier of the two states. To achieve a reliable switching of the Néel vector, it is preferable to generate a sufficiently large strain field that stabilizes only one type of domain. 

To understand how such strain arises, we consider an elastic medium subject to an inhomogeneous temperature distribution. It is intuitive that the lattice in the hotter region expands more significantly than in the colder regions. However, the expansion of the hot region creates an additional stress field that enhances the local deformation in its neighboring areas. As a consequence, the lattice displacement typically reaches its maximum at an intermediate temperature region, as illustrated in Fig. \ref{fig2}(b). For a predominantly horizontal (vertical) temperature gradient, the strain component $u_{xx}$ ($u_{yy}$), defined as the derivative of the displacement $u$, dominates and therefore determines the sign of $u_{me}$, which initially takes positive (negative) values and then reverses its sign to negative (positive) values \cite{Endmater}. In the following, we show that laser-induced plasmonic excitations can control the spatial distribution of temperature and, consequently, the strain field, laying the foundation of magnetic switching.

\begin{figure}
	\centering
	\includegraphics[width=0.48\textwidth]{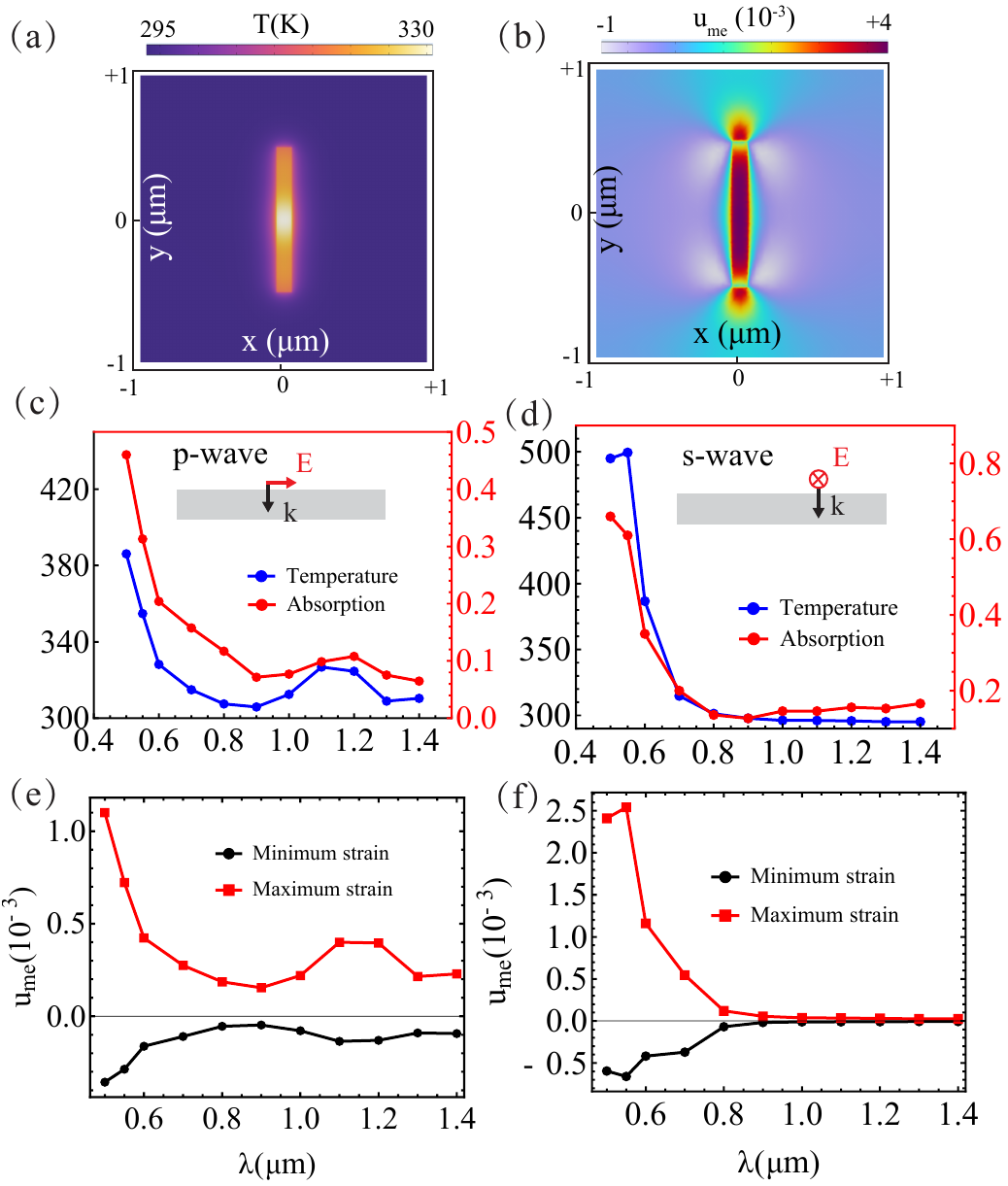}\\
	\caption{(a-b) Temperature distribution and the corresponding strain profile in a hybrid structure of gold nanowire$|$NiO thin film. The nanowire is cuboid-shaped with dimensions $0.1\times0.1\times1.0~\mathrm{\mu m}^3$ and the surface of the NiO film is $2 \times 2~\mu \mathrm{m}^2$. A single rectangular laser pulse (p-wave) with wavelength 1.2 $\mathrm{\mu m}$, duration 1 ns, and peak power 0.4 W incident perpendicularly on the hybrid structure. The temperature and strain profile were recorded at 1 ns. All the thermal and optical parameters of Au and NiO, as well as the simulation method, can be found in the End Matter. (c-d) shows the average temperature (blue dots) and absorption cross section of the gold nanowire (red dots) as a function of the wavelength of incident light for s-wave and p-wave configurations, respectively. (e-f) Wavelength dependence of the maximum and minimum strain field $u_{me}$ on the NiO surface.}\label{fig3}
\end{figure}

\textit{Plasmon-induced heating and strain.--} When a laser illuminates a metallic nanowire, the free electrons within the material are collectively excited, forming surface plasmons. The electromagnetic energy dissipated by these oscillations is converted into heat, raising the temperature of the metal. Then the heat diffuses from the hot nanowire into the surrounding medium, generating spatial temperature gradient and, consequently, thermal strains in the magnetic substrate.

To illustrate this mechanism, we first examine a cuboid-shaped gold nanowire patterned on an AFM insulator (NiO), as shown in Fig. \ref{fig3}(a). By numerically solving the Maxwell equations, the heat diffusion equation, and the thermal elastic equation using the COMSOL Multiphysics package \cite{comsol,Endmater}, we obtain the temperature distribution as well as thermal-induced strains in the hybrid structure.
Figure \ref{fig3}(a) shows the temperature rise in the gold nanowire under a normally incident laser pulse, while Fig. \ref{fig3}(b) displays the resulting thermal-induced strain. As anticipated from Fig. \ref{fig2}(b), the strain field is negative (positive) on the left/right (top/bottom) sides of the nanowire. Owing to the geometric anisotropy, the strain field also inherits a pronounced anisotropy, extending more broadly along the long edges of the nanowire. 

More importantly, this anisotropic strain distribution turns out to be highly tunable. By tailoring the nanowire aspect ratio and the polarization of the incident light, one can selectively excite distinct plasmon modes and thereby engineer the spatial pattern and magnitude of the strain field. In general, a metallic nanowire supports two characteristic plasmon modes: a longitudinal mode associated with electron oscillations along the long axis of the wire, and a transverse mode corresponding to oscillations along its short axis \cite{Baffou2013}. Their excitation efficiency depends strongly on the polarization of the incident wave. Because the longitudinal and transverse plasmon modes lead to very different spatial distributions of temperature, the induced strain field can thus be harnessed to manipulate the magnetic domain orientation. Specifically, for a p-polarized incident wave, the electric field oscillates along the longitudinal direction of the nanowire, mainly exciting the longitudinal plasmon modes. This is evident by a clear absorption peak and temperature increase around $\lambda=1.1~\mu \mathrm{m}$, where a distinct local resonance peak in both absorption and temperature rise appears, corresponding to the excitation of the longitudinal plasmon mode, as shown in Fig. \ref{fig3}(c). The resulting thermal-induced strain effect is also maximized at this wavelength (see Fig. \ref{fig3}(e)). The analytical calculation of the resonance position is challenging because the wavelength of the incident wave becomes comparable with the length of the nanowire, producing a complicated electric field distribution inside the nanowire. An intuitive interpretation follows from viewing the nanowire as a one-dimensional Fabry-P\'{e}rot resonator, guiding the surface plasmons along the wire \cite{DorfNL2009,AlbertAN2011}. Resonance emerges when the nanowire length ($1\mu \mathrm{m}$ here) approximately matches an integer multiple of the incident wavelength.

In contrast, for s-polarized illumination, the electric-field component of the EM wave oscillates along the thickness and width directions of the nanowire and mainly excites the transverse plasmon mode. A feature around $\lambda=0.55~\mu \mathrm{m}$, as shown in Fig. \ref{fig3}(d), is consistent with the expected transverse plasmon response of metallic nanowires with comparable dimensions reported in literature \cite{Baffou2010, Bohren1983} and with Mie-type considerations. For the long wavelength $\lambda=1.1~\mu \mathrm{m}$, both the temperature rise and the thermal induced-strain are greatly reduced. This strong contrast between the resonant responses for p- and s-polarizations demonstrates that plasmon excitation and therefore the thermal strain field can be selectively enhanced or suppressed simply by changing light polarization. Such selectivity forms the foundation for reversible magnetic switching discussed below.

\textit{Reversible switching inside the metallic frame.--} We now demonstrate reversible AFM switching in a hybrid gold square frame and AFM insulator structure, as illustrated in Fig. \ref{fig1}(a). The mechanism follows directly from the polarization-selective excitation of the plasmon modes supported by the frame arms. When an s-wave with a frequency near the longitudinal plasmon resonance illuminates the structure, the electrons in the vertical arms of the square are primarily excited. The resulting strain field inside the frame is dominated by heating in the two vertical arms and is typically negative ($u_{me}<0$) inside the square, as shown in Fig. \ref{fig3}. In contrast, when a p-wave near the longitudinal resonance frequency is incident, plasmon excitation in the vertical arms is suppressed because the electric field of the incident wave is perpendicular to the longitudinal direction. Instead, the two horizontal arms will be excited and the resulting thermal induced strain field $u_{me}$ reverses its sign, i.e., $u_{me}>0$. Consequently, the Néel vector inside the square will be switched from $n_x=\pm 1$ to $n_y=\pm 1$ reversibly as the polarization of the incident wave is alternated.

\begin{figure}
	\centering
	\includegraphics[width=0.48\textwidth]{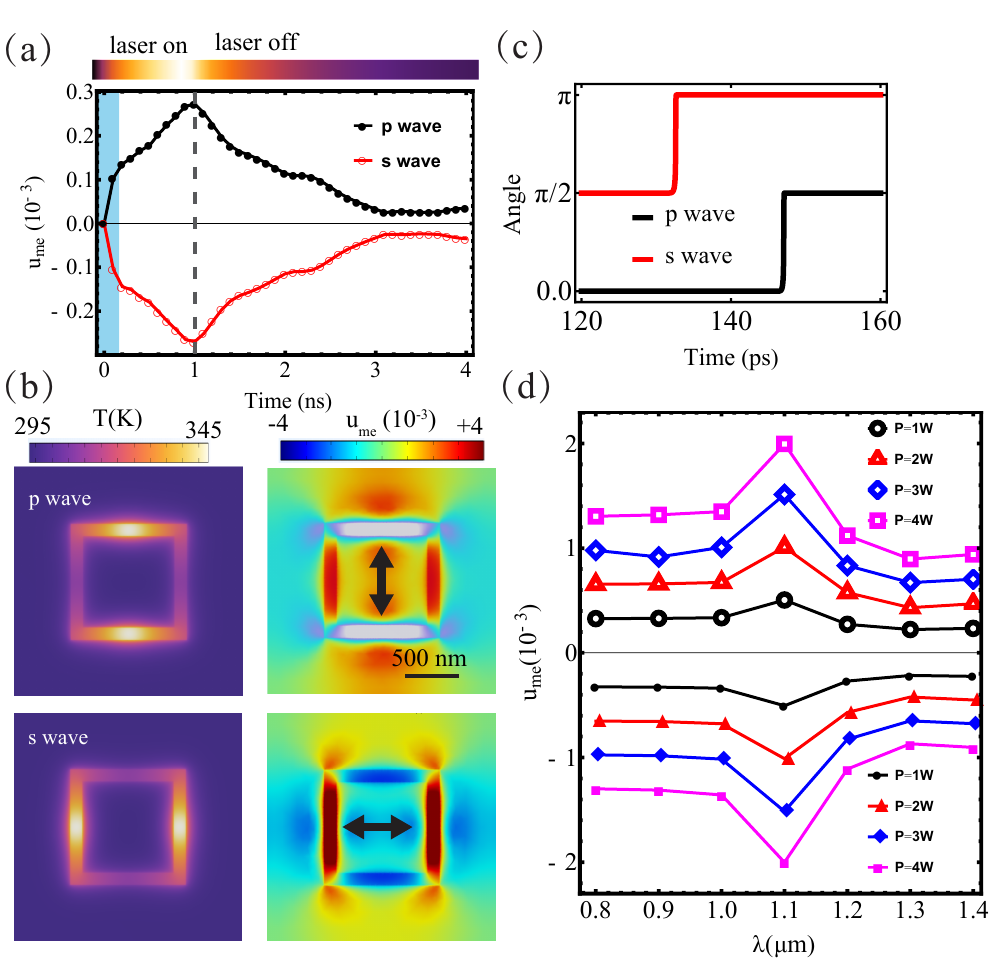}\\
	\caption{(a) Time evolution of the surface strain field driven by a 1 ns laser pulse in a hybrid structure of gold $\mathrm{nanowire}|\mathrm{NiO}$ thin film. The colored rectangle represents the temperature of the nanoframe at the corresponding time, with the same color bar shown in Fig. \ref{fig4}(c). $\lambda=1.2~\mu \mathrm{m}, P =1~\mathrm{W}$. Geometric parameters are the same as those in Fig. \ref{fig3}. (b) Snapshot of temperature profile and the corresponding strain distribution at the magnetic surface at $t=1~ \mathrm{ns}$. (c) Time evolution of the in-plane tilting angle of the Néel vector under the incidence of p wave (black line) and s wave (red line), respectively. (d) Strain field as a function of the wavelength of incident light for s-wave (solid symbols) and p-wave (open symbols) configuration, respectively.}\label{fig4}
\end{figure}

The temporal evolution of the strain field for both polarizations is presented in Figure \ref{fig4}(a), where solid (open) symbols correspond to p-(s-) polarized excitation and the accompanying colored rectangular bar indicates the average temperature of the nanoframe. When the laser is turned on at $t=0$ ns, both the temperature of the nanowire and the resulting strain field within the frame rapidly increase and reach their maxima at $t\approx 1$ ns for the p-wave. After the laser is turned off at $t=1$ ns, both quantities decay gradually as the structure cools. Representative snapshots of temperature and strain field distribution are plotted in Fig. \ref{fig4}(b). For the p-wave polarization, it is evident that only the two horizontal arms are excited, with two distinct hot spots clearly visible. The average strain field reaches $u_{me}=0.72 \times 10^{-4}$, which exceeds the critical threshold required to reorient the Néel vector along the $y-$axis. The switching happens on an ultrafast timescale once the strain field surpasses the threshold, as shown in Fig. \ref{fig4}(c). 

Reversibility is achieved by switching back to s-polarized illumination, which excites the two vertical arms and produces a negative strain field that restores the Néel vector to the $x-$axis. Figure \ref{fig4}(d) presents the wavelength dependence of the maximum and minimum strain fields at different laser powers. For a given power, the strain field exhibits a peak (for s-waves) or a valley (for p-waves) around $1.1~\mu \mathrm{m}$, where the longitudinal plasmon mode of the vertical (horizontal) arms is resonantly excited. At all wavelengths, the strain amplitude increases linearly with the laser power, confirming that the thermal effects dominate the strain generation. These results establish that polarization-selective plasmon excitation provides a robust and fully reversible means for controlling the Néel-vector orientation inside the nanoframe.

\textit{Discussion and conclusions.}--- The plasmon-induced magnetic switching is governed by a sequence of distinct timescales. Immediately after optical excitation, hot electrons are generated in the metallic nanostructure on an ultrafast timescale of 1-100 fs. This is followed by electron-electron scattering (approximately 1 ps), which redistributes their energy and elevates the temperature of nanostructure \cite{Jia2016}. On a longer timescale of 0.1-1 ns, the localized thermal energy is transferred to the underlying AFM layer, giving rise to inhomogeneous thermal strains. Since the Néel vector responds on the intrinsic picosecond timescale of AFM dynamics, it rapidly reorients to follow the direction of the strain field. This separation of timescales ensures that the full switching cycle is primarily governed by the slower evolution of the thermal landscape on nanosecond timescale.

%\textcolor{red}{Compared with ferromagnets, antiferromagnets Our findings provide an all-optical approach to manipulate the antiferromagnets, 

\begin{table}[h!]
\centering
\caption{Comparison among various switching schemes of antiferromagnets. ec and op are short for electric current and optical pulses, respectively. }
\begin{tabular}{l|c|l|c|c}
    \hline
    \hline
  \shortstack{Hybrid\\ structure} & \shortstack{Rotating\\ angle}&\shortstack{Pulse \\ width(ms)} &\shortstack{Pulse amp\\ ($\mathrm{mA}$)} &\shortstack{Switching \\ energy ($\mu J$)} \\
  \hline
   CuMnAs \cite{Wadley2016}& $90^\circ$ & 50 (ec) & 90 &$8100$\\
   NiO/Pt \cite{Meernl2021}&$90^\circ$ & 0.1 (ec) & 7 & $3$ \\
  $\mathrm{Fe_2O_3}$/Pt \cite{Zhangprl2019} &$90^\circ$ & 10 (ec)  & 40 & $4528$\\
 $\mathrm{Mn_3Sn}$/Pt \cite{Tsai2020} &$180^\circ$ & 100 (ec) & 30&  $3452$\\
  CoO/Pt \cite{Wuprb2024}& $90^\circ,180^\circ$ &0.1 (ec)& 15 &$8$\\
  $\mathrm{Mn_5Si_3/Pt}$ \cite{HanSA2024} &$180^\circ$& 1 (ec) &46 & 89 \\
  \hline
 \textbf{NiO/Au} &$\mathbf{90^\circ}$&\textbf{$\mathbf{10^{-6}}$(op)}& \textbf{NA} &\textbf{0.001}\\
\hline
\end{tabular}
\label{tab1}
\end{table}

To contextualize the energy efficiency of our approach, we compare the switching energy with that reported in current-driven AFM switching experiments. In those systems, the switching energy can be estimated as $\Delta E= I^2 R\Delta t$, where $I, R, \Delta t$ are respectively the switching current, device resistance and duration of the current pulse. In our proposal, the switching energy is $\Delta E=P\Delta t$. As summarized in Table \ref{tab1}, the switching energy via optical plasmons is on the order of  $0.001 \mu J$, corresponding to a reduction of three to six orders of magnitude compared to current-driven techniques. Such a dramatic improvement likely arises from a combination of factors including the limited charge-to-spin efficiency in electric schemes, the long current pulses typically required and the crossbar geometry making the electrical path highly dissipative. In contrast, the strong field localization of plasmons ensures that nanoscale heating is generated precisely near the frame arms, allowing the system to harvest thermal energy with minimal dissipation. The endurance of the proposed scheme may also be favorable compared with current-driven approaches, since the switching process does not involve significant Joule heating.
In addition, we emphasize that the present geometry is chosen for demonstrating the essential physics rather than for device optimization. The thermal diffusion length $L_\mathrm{th}\sim \sqrt{\alpha \Delta t}$, where $\alpha$ is the thermal diffusivity, sets a relevant scale for heat localization rather than a strict constraint on the device geometry. Using typical values of $\alpha \sim 10^{-5}~\mathrm{m^2/s}$ \cite{Endmater} and $\Delta t=1~\mathrm{ns}$, $L_\mathrm{th}$ is on the order of 100 nm. This suggests that the sidelength of proposed geometry may be scaled down to the sub-100 nm regime without altering the underlying switching mechanism.

Finally, we examine the robustness of the mechanism against temperature-dependent parameters. In our simulations, the temperature rise in the magnetic layer during switching remains below 50 K. Experimental measurements show that the thermal expansion coefficient of NiO varies only weakly ($\sim 1\%$) in this temperature range \cite{Nielsen1965}, corresponding to a density variation of only $0.2\%$. The heat capacity increases by roughly $5\%$ \cite{Lewis1973}, leading to only a minor modification of the thermal response. The magnetic parameters entering Eq. \eqref{eqn1} are also expected to vary weakly for such a moderate temperature change, since the operating temperature remains well below the Néel temperature of NiO ($T_N \approx 525$ K). Non-ideal lasers and environmental factors such as substrate coupling and modified heat-dissipation pathways may slightly alter the detailed heating profile and cooling dynamics, but they do not affect the underlying polarization-controlled switching mechanism. Consequently, the strain-induced switching mechanism is robust against reasonable temperature variations.

In summary, we have demonstrated that selective heating induced by surface plasmons can be harnessed to manipulate antiferromagnetism with ultralow energy consumption. The underlying physics is governed by thermal-induced and anisotropic strain fields, which couple to the Néel vector and drive switching between two orthogonal AFM domains. By selectively exciting the longitudinal and transverse plasmon modes of the nanoframe, the sign and magnitude of the strain field can be precisely controlled, enabling reversible switching of the Néel vector. The same principles may be extended to ferromagnets with four-fold symmetry through magnetoelastic coupling. More broadly, integrating plasmon-induced heating with magnetic platforms introduces a versatile framework for exploring a wide range of spintronic phenomena, including spin caloritronics, magnetic-texture-based spintronics, and hybrid magnonics. Furthermore, our proposed architecture is intrinsically energy-efficient and scalable, where each frame in an array can be independently controlled by tailored optical pulses. These features position our approach as a promising platform for high-density, all-optical magnetic storage and processing.
%{\color{red} Do we know anything about the polarization of the excited plasmons? If yes, and if we can generate TE plasmons, we should say this here. If not, we might want to work a bit the intro because it clearly creates an impression that we are dealing with TE plasmons.}

{\it Acknowledgments.}---H.Y.Y acknowledges the helpful discussions with Tai Min and Weichao Yu. This work is supported by the National Key R$\&$D Program of China (2022YFA1402700 and 2024FYA1408500) and the National Natural Science Foundation of China (NSFC) (Grant No. 12574132 and 12221004). O.G. acknowledges support from the Deutsche Forschungsgemeinschaft (DFG, German Research Foundation) - TRR 173 – 268565370 (project A11) and TRR 288 – 422213477 (project A12).

%	\newpage
%	\bibliographystyle{apsrev4-2}
%	\bibliography{bibliographysot}
%\bibliography{magnon_plasmon}
{}

\section{End Matter}
\textit{Axially symmetric heating induced strain.--}
To illustrate the essential feature of thermal-induced strains, we consider a cylinder medium with an axially symmetric temperature distribution, then the strain field also has axial asymmetry, and the strain field in the radial direction $(u)$ follows the steady equation of motion \cite{Landaubook,Hetnarski}
\begin{equation}
\frac{d}{dr}\left ( \frac{1}{r} \frac{d(ru)}{dr}\right) = \frac{1+\nu}{1-\nu} \alpha \frac{dT(r)}{dr},
\end{equation}
where $\nu$ is Poisson's ratio, defined as the negative ratio of transverse strain to axial strain, and $\alpha$ is the thermal expansion coefficient of the medium. Considering the constraint that the thermal stress is zero at the sample boundary $r=R$, we can analytically solve the strain field as
\begin{equation}
u=\frac{1+\nu}{1-\nu}\alpha \left [ \frac{1}{r} \int_0^r rTdr + (1-2\nu) \frac{r}{R^2}\int_0^R Trdr\right].
\end{equation}
Take the example of an exponentially decay temperature profile $T=T_0 e^{-r^2/a^2}$ with $T_0$ being the center temperature and $a$ being the decay distance of the temperature field (see Fig. \ref{fig2}(a)), we can derive the strain field around the cylinder center line as 
\begin{equation}
u=\alpha T_0 \frac{a^2(1+\nu)}{2(1-\nu)} \frac{(1-e^{-r^2/a^2})}{r^2}.
\end{equation}
In the Cartesian coordinate, the strain tensor components are $u_{xx}=\cos^2\theta \partial_ru$ and $u_{yy}=\sin^2\theta \partial_ru$, from which the effective magnetoelastic field acting on the magnetic order is evaluated as $u_{me}= u_{xx}-u_{yy}=\cos 2\theta \partial_ru$, i.e.
\begin{equation}
u_{me}=\alpha T_0 \frac{1+\nu}{1-\nu} \cos 2\theta \left [ \left(1+\frac{a^2}{2r^2}\right)e^{-r^2/a^2}-\frac{a^2}{2r^2} \right ].
\end{equation}

\begin{figure}
	\centering
	\includegraphics[width=0.48\textwidth]{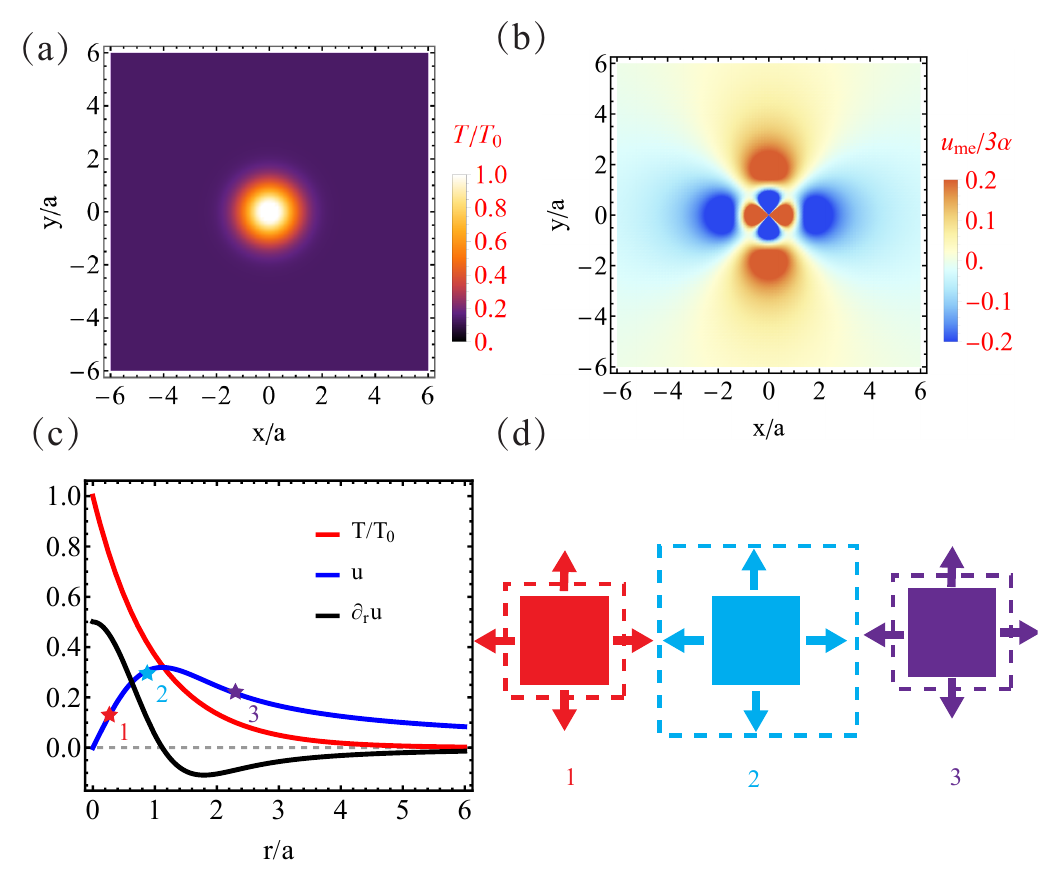}\\
	\caption{(a-b) Axially symmetric temperature profile and the subsequent thermal-induced strain field distribution in an elastic medium. $\nu=1/2$ is used. (c) Radial dependence of the temperature (red line), radial displacement (blue line), and strain tensor component (black line). (d) Illustration of the lattice deformation in a medium under an inhomogeneous temperature distribution.}\label{figem}
\end{figure}

Figure \ref{figem}(b) shows the typical spatial profile of $u_{me}$, which takes on a rotational asymmetry due to the angular dependence $\cos 2\theta$. Let us first examine the strain field along the $x-$direction. Intuitively, the lattice in the sample center $r=0$ with higher temperature expands much more severely than the outside region with lower temperature. However, this trend will be suppressed at a certain distance since the lattice outside will also be squeezed by the expansion of the inner lattice, resulting in a maximum of the strain field in the middle part of a sample, as shown in Fig. \ref{fig2}(c) (blue line). Then the strain tensor $u_{xx}$ defined as the derivative of the displacement $u$ first takes on a positive value and then reverses its sign to a negative value. This explains the strain field profile along the $x-$direction, and a similar argument explains the strain profile along the $y-$direction. 

\textit{Methodology.--} In most cases, the thermal-induced strain distribution is not analytically solvable. The following steps are followed to simulate the optical response of the hybrid structure we considered. Firstly, a hybrid Au/NiO structure was constructed and discretized using the finite element method in COMSOL with physics-controlled mesh. Secondly, the Maxwell equation is solved to obtain the spatial distribution of electromagnetic fields upon the incidence of a laser using the \textit{Optics} module. Thirdly, based on the calculated electromagnetic distribution, the electromagnetic
loss rate $q(r)=\epsilon_0 \Im (\epsilon(\omega)\mathbf{E}(\mathbf{r})\mathbf{E}^*(\mathbf{r}))/2$ is calculated and modelled as a heat source of the hybrid structures. Then, the heat transfer process with thermal contact between the Au nanowire and magnetic film is calculated by numerically solving the thermal diffusion equation using the \textit{Heat Transfer in Solids} module. Fourthly, the strain distribution $u_{xx}$ and $u_{yy}$ in the magnetic film is obtained by solving the thermal-elastic equation using the \textit{Solid Mechanics} module. The strain field is calculated as $u_{me} = u_{xx} - u_{yy}$. Fifthly, the strain field is then fed into a home-made micromagnetic simulation code to simulate the dynamics of Néel vector governed by the equation
\begin{equation}
\mathbf{n} \times (\alpha \partial_t \mathbf{n} - A \nabla^2 \mathbf{n} + \mathbf{H}_\mathrm{an})=0
\end{equation}
where $\alpha$ is damping parameter, $A$ is exchange stiffness, $\mathbf{H}_{an}= -1/(2M_s) \delta \mathcal{H}/\delta \mathbf{n}$ is the effective anisotropy field including the contribution from intrinsic anisotropy and magnetoelastic effect.

We used built-in parameters of Au and NiO in COMSOL if available, as indicated in Table \ref{tab2} below. The elastic parameters of NiO are Young's modulus $E=1.718 \times 10^{11}$ Pa, Poisson's ratio $\nu=0.348$ \cite{Meernl2021}. The magnetic parameters are $H_\| = 1$ T, $H_\perp = 0.1$ T, Gilbert damping $\alpha=0.01$. The thermal diffusivity of NiO can be calculated as $\alpha =k/(\rho c_p) =1.3 \times 10^{-5} \mathrm{m^2/s}$ with $k, \rho, c_p$ being the thermal conductivity, density, and heat capacity, respectively. 

\begin{table}[h!]
\centering
\caption{Parameters used in the simulations.}
\begin{tabular}{l|c|c}
    \hline
    \hline
  Material & Au &NiO  \\
  \hline
   Density ($\mathrm{g/m^3}$) &$19.3$&6.67 \\
   Thermal conductivity ($\mathrm{W/(m\cdot K)}$) & 317 &50\\
   Heat capacity ($\mathrm{J/(kg\cdot K)}$) & 129 &559\\
   Thermal expansion ($\mathrm{1/K}$) &NA&$1.41\times 10^{-5}$\\
\hline
\end{tabular}
\label{tab2}
\end{table}
\end{document}